\documentclass{emulateapj}
\usepackage{longtable}

\def\lap{\lower.5ex\hbox{$\; \buildrel < \over \sim \;$}}
\def\gap{\lower.5ex\hbox{$\; \buildrel > \over \sim \;$}}

\def\ergcm2s{${\rm erg\ cm^{-2}\ s^{-1}}$}

\def\ergscm2s{${\rm erg\ cm^{-2}\  s^{-1}}$}

\def\cm-2{${\rm cm^{-2}}$}

\def\pc{{\rm{\,pc}}}

\def\Mpc{{\rm{\,Mpc}}}
\def\Myr{{\rm{\,Myr}}}

\def\mzams{M$_{\rm ZAMS}$}

\begin{document}

\title{Constraints for the Progenitor Masses of 17 Historic Core-Collapse Supernovae}

\author{Benjamin F. Williams\altaffilmark{1},
Skyler Peterson\altaffilmark{1},
Jeremiah Murphy\altaffilmark{2},
Karoline Gilbert\altaffilmark{1,3},
Julianne J. Dalcanton\altaffilmark{1},
Andrew E. Dolphin\altaffilmark{4},
Zachary G. Jennings\altaffilmark{5}
}

\altaffiltext{1}{Department of Astronomy, Box 351580, University of Washington
, Seattle, WA 98195; ben@astro.washington.edu; peters8@uw.edu; jd@astro.washington.edu}
\altaffiltext{2}{Department of Physics, Florida State University, jeremiah@physics.fsu.edu}
\altaffiltext{3}{Space Telescope Science Institute; kgilbert@stsci.edu}
\altaffiltext{4}{Raytheon, 1151 E. Hermans Road, Tucson, AZ 85706; dolphin@raytheon.com}
\altaffiltext{5}{University of California Observatories, Santa Cruz, CA 95064; zgjennin@ucsc.edu}


\begin{abstract}

Using resolved stellar photometry measured from archival {\it HST} imaging, we generate color-magnitude diagrams of the stars within 50 pc of the locations of historic core-collapse supernovae that took place in galaxies within 8 Mpc.  We fit these color-magnitude distributions with stellar evolution models to determine the best-fit age distribution of the young population.  We then translate these age distributions into probability distributions for the progenitor mass of each SNe.  The measurements are anchored by the main-sequence stars surrounding the event, making them less sensitive to assumptions about binarity, post-main-sequence evolution, or circumstellar dust.  We demonstrate that, in cases where the literature contains masses that have been measured from direct imaging, our measurements are consistent with (but less precise than) these measurements. Using this technique, we constrain the progenitor masses of 17 historic SNe, 11 of which have no previous estimates from direct imaging.  Our measurements still allow the possibility that all SNe progenitor masses are $<$20 M$_{\odot}$.  However, the large uncertainties for the highest-mass progenitors also allow the possibility of no upper-mass cutoff.

\end{abstract}

\section{Introduction}

A fundamental question of stellar evolution theory is which stars end
their lives as supernovae.  Current theory for isolated massive stars
makes two basic predictions for SNe.  For one, the
zero-age-main-sequence mass (\mzams) and mass-loss history control
whether a SN occurs, and secondly, for single stars there is a clear
mapping between \mzams\ and the type of SN
\citep{woosley02,heger03a,dessart11}.  In particular, the lower masses
explode with their hydrogen envelopes intact (e.g. II-P, II-L, IIn),
and the most massive stars lose much of their envelopes and explode as
hydrogen deficient SNe (e.g. IIb, Ib/c).  However, given the
complexity of the underlying physics, especially binary evolution,
winds, and episodic eruptions, it is unclear whether nature obeys the
same well-delineated mass-dependence.

In fact,the relatively high observed rates of H-deficient SNe
\citep{smith11c} and low upper limits on progenitor masses of Type Ibc
SNe \citep{yoon2012,eldridge2013} imply that binary evolution may
figure prominently in producing the H-deficient SNe.  Furthermore,
theory predicts that binary evolution can significantly affect the
mapping between initial stellar mass as SNe type
\citep[e.g.,][]{podsiadlowski1992,tutukov1992,nomoto1995,dedonder1998,yoon2010bin,claeys11,dessart11,eldridge2011}.
It is clear that progress in understanding both mass loss and SNe
requires observational constraints linking the progenitor mass with
the eventual SN.

Unfortunately, of the dozens of SNe that have progenitor mass limits,
only 17 have masses measured from a directly detected progenitor (6 of
these 17 overlap this work).  \citet{smartt09b} reviewed the mass
distribution of 30 core-collapse SNe progenitors (20 type IIP and 10
others), but only eight of these had measurements beyond an upper
limit.  At that point, there were also 4 other nearby SNe progenitors
with full mass constraints
\citep{woosley1988,aldering1994,crockett2008,fraser2010}, bringing the
total to 12.  Since 2009, 5 additional SNe have been measured
\citep{maund11,murphy11a,fraser2012,vandyk2012b,maund2013,fraser2013,vandyk2014b},
and some measurements improved \citep{vandyk2012a,maund2014} making
the total 17 measurements.

These mass estimates are based on serendipitous direct imaging of the
progenitor.  First, one searches through the HST archive for bright
evolved stars at the SN position.  Then, if a star is found, the
endpoints of stellar evolution models which pass through the color and
magnitude of the likely progenitor are used to estimate the star's
initial main sequence mass.  If a star is not found, an upper limit on
the progenitor luminosity is measured.

Even with the limited number of measurements available, there is a
hint of a minimum \mzams\ for explosion and that the least massive
stars explode as SN II-P \citep{smartt09a}. Interestingly, these
measurements also suggested that the maximum mass for SN II-P may be
lower than expected, with some perhaps having been merged binaries
\citep{smartt09a}, although circumstellar dust may also explain the
observations \citep{walmswell2012}.  Counter to expectations, some
H-rich SNe (in particular IIn) have been associated with very massive
stars \citep{galyam09,smith11b,smith11c}. While tantalizing, these
initial results are poorly constrained, and even the simple
$\sim$8~M$_{\odot}$ lower mass limit requires more observational
constraints.

While direct imaging of progenitors is the standard method for
progenitor mass estimation, it suffers from a number of limitations.
First among these is the requirement that the precursor imaging
actually exist. The majority of past SNe have neither pre-existing HST
imaging, nor sufficiently accurate astrometry. Consequently, of the
$\sim\!40$ historic SNe within $\sim\!10$ Mpc, only a handful have
identified progenitors.  Some future nearby SNe may also lack
precursor imaging due to the limited observations in the HST archive.

The second major limitation is that even when precursor imaging is
available, interpretation of that imaging depends on modeling of the
most uncertain stages of stellar evolution
\citep{gallart05,smartt09a,yoon2010,langer2012,eldridge2013}.
Existing studies estimate the mass of a precursor by fitting endpoints
of stellar evolution models to its color and magnitude; however, an
evolved star's appearance is not well constrained during the final
evolutionary stages.  Binarity, mass loss, pulsation, internal mixing,
the formation of dust in stellar winds, and convective instabilities
in shell-burning layers all contribute to systematic and random
uncertainties in such model endpoints.  Matching individual endpoints
of stellar evolution models to a single highly-evolved star on the
brink of explosion therefore places weak constraints on the stellar
mass once systematic uncertainties are taken into account.

In this paper, we take a complementary approach that obviates both of
these limitations.  We build on a technique developed by several
investigators over the past two decades, starting with measuring ages
for host star clusters
\citep{efremov1991,walborn1993,panagia2000,barth1996,vandyk1999,maiz2004,wang2005,vinko2009,crockett2008},
moving into finding coeval field populations around both SNe and
supernova remnants
\citet[SNRs;][]{badenes2009,gogarten2009,murphy11a,jennings2012}. We
note that, while we have applied this approach to many SNR locations,
historic SNe are less numerous, but more reliable, targets because
they have well-established locations and types, ensuring that every
progenitor mass corresponds to a {\it bona~fide} core-collapse event.

The technique finds the masses of SNe precursors by analyzing the
stellar populations of stars surrounding the SNe.  Using
well-established techniques of stellar population modeling, we can
age-date the star formation (SF) event that led to each SN.  The
resulting age places strong constraints on the mass of the precursor,
using the well-understood properties of main-sequence stars. As a
result, in cases where the specific stellar population can be
identified, our technique provides a more reliable progenitor age than
direct imaging, as it does not depend on whether the progenitor was a
single star or a binary.  Furthermore, the method works even when
there is no imaging prior to the SN, or when the SN position is only
localized to within a few arcseconds.  Hence our technique can be
applied to the location of any historic SN that has sufficiently
high-resolution and deep imaging to measure resolved stellar
photometry of the upper main sequence and/or He-burning sequence.

In \S~2, we discuss our sample, the data analyzed in our study, detail
our analysis technique, and demonstrate its efficacy in test cases.
In \S3, we provide our results in the form of age distributions, a
table of masses, and a table of probability distributions for each
progenitor.  Finally, \S~4 gives a summary of our findings.

\section{Data}

\subsection{Sample}

We selected all positively-typed historic core-collapse SNe within $8
\Mpc$ that also have $\sim$arcsecond accuracy in their positions from
the Asiago Supernova Catalog \citep{barbon1999}; higher positional
accuracy is not needed, due to the large size of the aperture within
which the CMD is constructed.\footnote{One arcsec corresponds to 5 pc
  for every Mpc of distance.}  We have shown in \citet{murphy11a} that
our method provides results consistent with direct progenitor
detections \citep[as confirmed by][]{vandyk2013} in galaxies as
distant as 8~Mpc. Beyond this limit, our method is not tested, and
therefore we have confined our sample to be within this distance.

We cross-referenced the SNe catalog with the HST archive and
identified SNe that have HST imaging in at least 2 broadband filters
in ACS, WFPC2, or UVIS.  Even relatively shallow data can provide
constraints given that, for the most massive $\sim$50~M$_{\odot}$
progenitors, the surrounding populations are likely to have other very
massive stars with M$_{\rm V}{<}{-}$5.  We found 22 SNe that match
these requirements.  Another SN, 2011dh, has already been analyzed by
our technique in \citet{murphy11a}.  Of our sample there are two that
are possibly SN impostors.  These are SN1954J \citep{smith2001} and
SN2002kg \citep{weis2005}.  We still include these impostors because
the classification as SN impostors are not definitive and it is likely
that these transients are associated with the last stages of stellar
evolution, in which our proposed method to derive progenitor masses is
still of interest. Table~\ref{sample} shows these SNe for which we can
attempt to derive the SFH and progenitor mass, along with the proposal
ID for the dataset used for out photometry.

There were five SNe which we attempted to analyze, but were not able to
constrain with confidence.  These had relatively shallow data for
quite distant events (SN~1980K, SN1985F, SN2002ap, SN2002bu, and
SN2003gd) with fewer than 5 stars detected within a 50 pc physical
radius. We do not consider these due to the sparsity of data, although
it is likely that deeper imaging of these locations would yield
photometry that would result in reliable mass estimates. However it is
also possible these progenitors were runaway stars exploding some
distance from their co-eval population \citep[c.f.,][]{eldridge2011}.

\subsection{Analysis Method}

Our method has been described in several other publications, including
its application to SN 2011dh \citep{murphy11a}, 121 SN remnants (SNRs)
in M31 and M33 \citep{jennings2012,zach2}, and an unusual transient in
NGC~300 \citep{gogarten09a}. We provide a description of the method
here as well for convenience.

In brief, we fit SFHs to CMDs of the population surrounding the site
of the SN to determine the age, and thus mass, of the SN progenitor.
The measurements are anchored by the main-sequence stars surrounding
the event; thus, our age estimates do not depend on whether a binary
or single-star progenitor is assumed.  Furthermore, the measurements
are not sensitive to any circumstellar dust present around the
progenitor itself.

In the following subsections we first summarize how well the method
works and provide a proof of concept.  Then we detail how our
photometry was performed, how the stellar samples were generated, and
how their age distributions were derived.

\subsubsection{Overview}

The method takes advantage of the fact that most stars form in stellar
clusters \citep{lada03} with a common age ($\Delta
t\!\lesssim\!1-4\Myr$) and metallicity.  Indeed, over 90\% of stars
form in rich clusters containing more than 100 members with
$M\!>\!50$M$_{\odot}$ \citep{lada03}.  The stars that formed in a
common event remain spatially correlated on physical scales up to
$\sim\!100\pc$ during the $100\Myr$ lifetimes of $4$M$_{\odot}$ stars,
even if the cluster is not gravitationally bound \citep{bastian06}; we
have confirmed this expectation empirically in several test cases
\citep{gogarten09a,murphy11a}.  Thus, it is reasonable to assume that
most young stars within a 50 parsecs of many SN are coeval.  However,
we note that our assumption breaks down for SNe from runaway stars
\citep{eldridge2011}, which would not be coeval with their surrounding
population.

The age of a SN's host stellar population can be recovered from its
color-magnitude diagram (CMD).  In the simplest method, one can fit a
single isochrone to an observed CMD and estimate the turnoff mass of
the youngest stars.  However, due to the small numbers of massive
stars, one can easily underestimate the mass, since CMDs can show an
apparent turnoff that is fainter than the true turnoff luminosity
simply because of poor Poisson sampling of the upper end of the IMF.
Instead, we adopt more sophisticated methods that take advantage of
the entire CMD.  These methods fit superpositions of stellar
populations to reproduce the observed CMD, using the recovery of
artificial stars to generate realistic distributions of stars from
theoretical isochrones.  The recovered recent SFH therefore fits not
just the turnoff luminosity, but the full luminosity function of the main
sequence and the blue and red core Helium-burning sequences as well.
Including the well-populated lower end of the main sequence adds
significant statistical weight when interpreting the sparsely sampled
population of massive upper main sequence stars.

The method allows dust extinction to be reliably taken into account.
The main sequence has a well defined color, such that any shift
towards redder colors must be produced by foreground reddening,
allowing the dust extinction to be inferred from the CMD itself.
Differential extinction can be constrained as well, using the observed
widening of the main sequence over what is expected from photometric
errors.  The resulting reddening constraints are dominated by the
young stars in which we are most interested.

\subsubsection{Method Validation}

\begin{figure}
\includegraphics[width=3.3in]{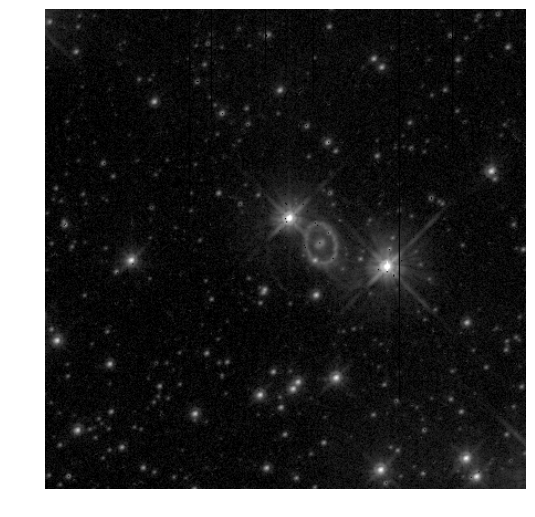}
\includegraphics[width=3.3in]{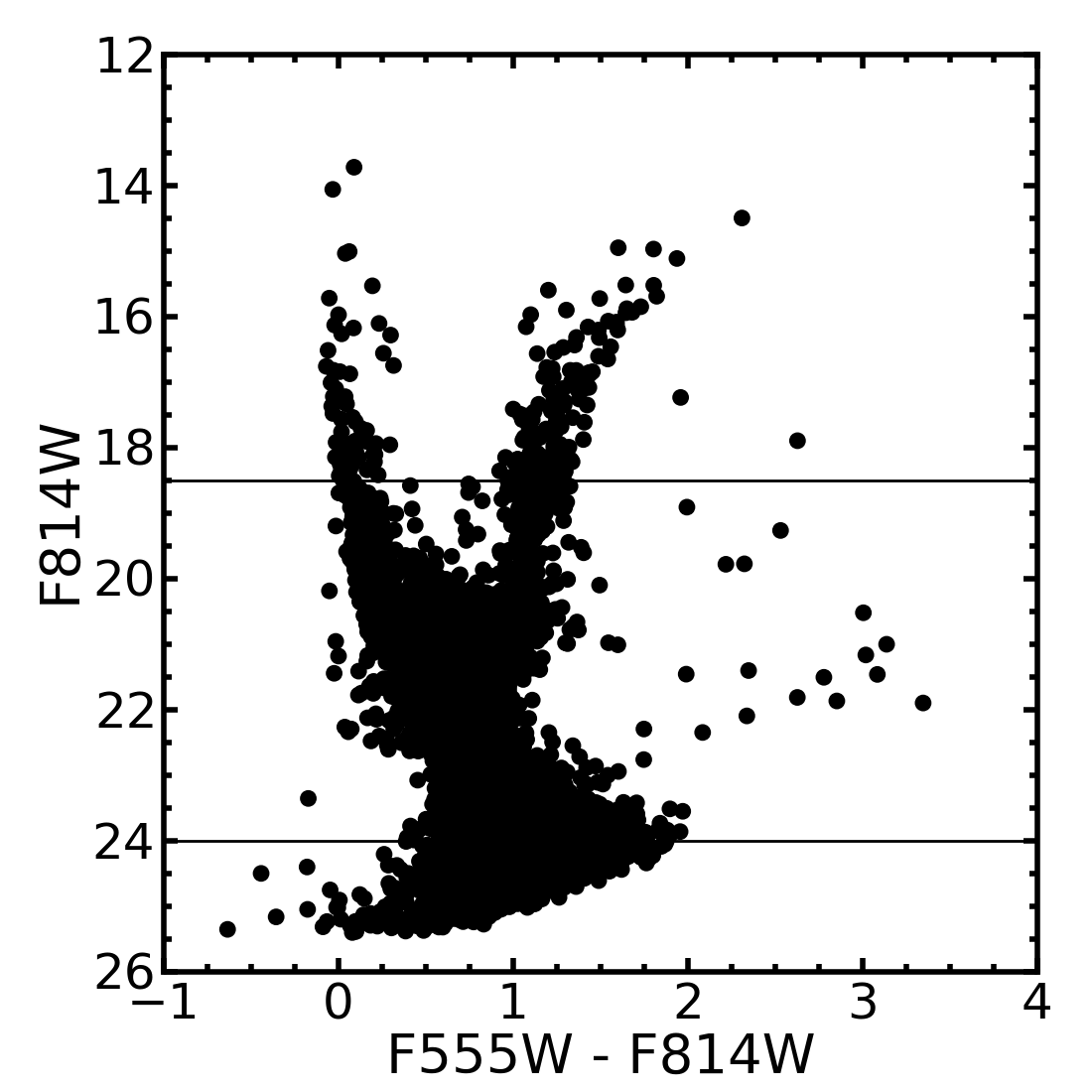}
\includegraphics[width=3.3in]{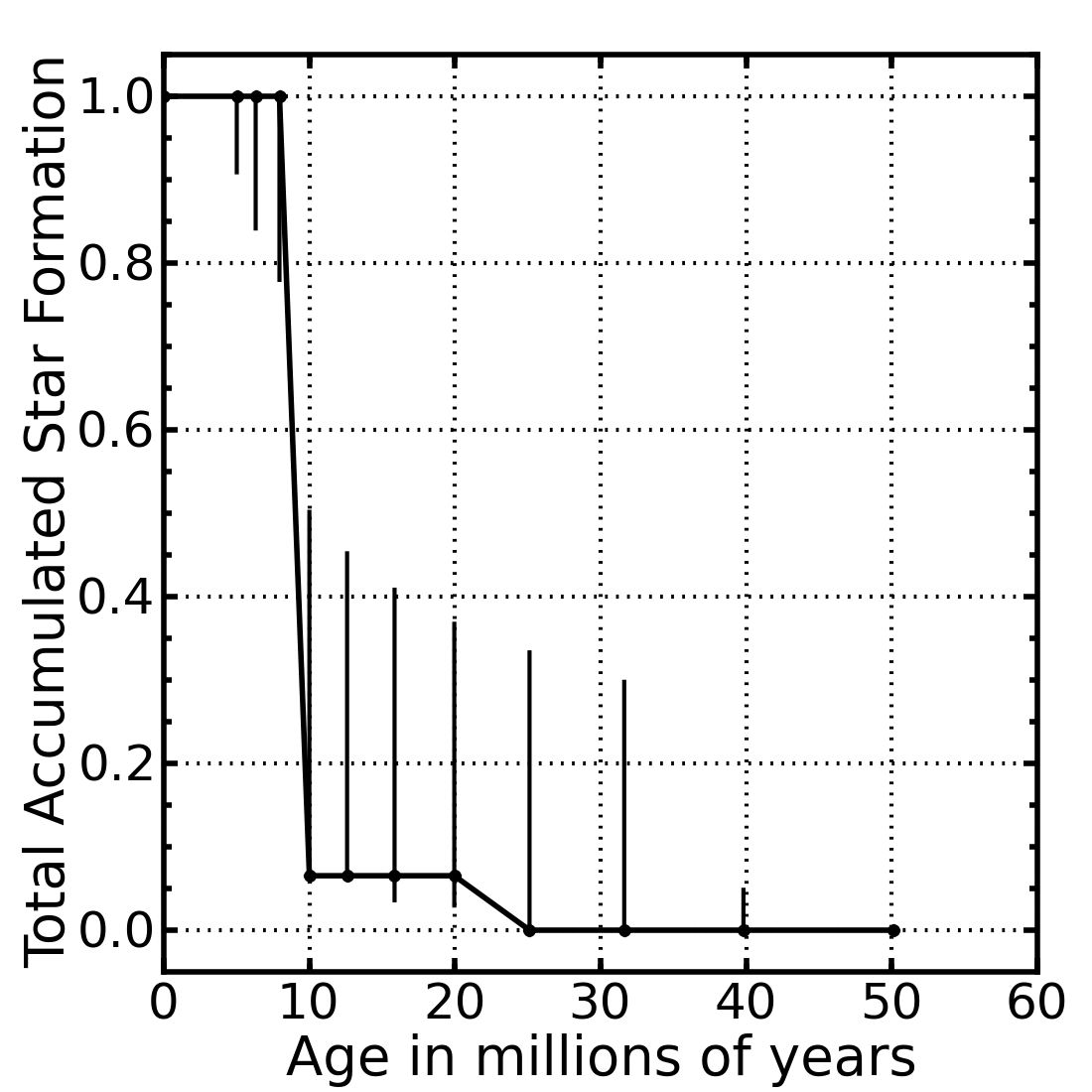}
\includegraphics[width=3.3in]{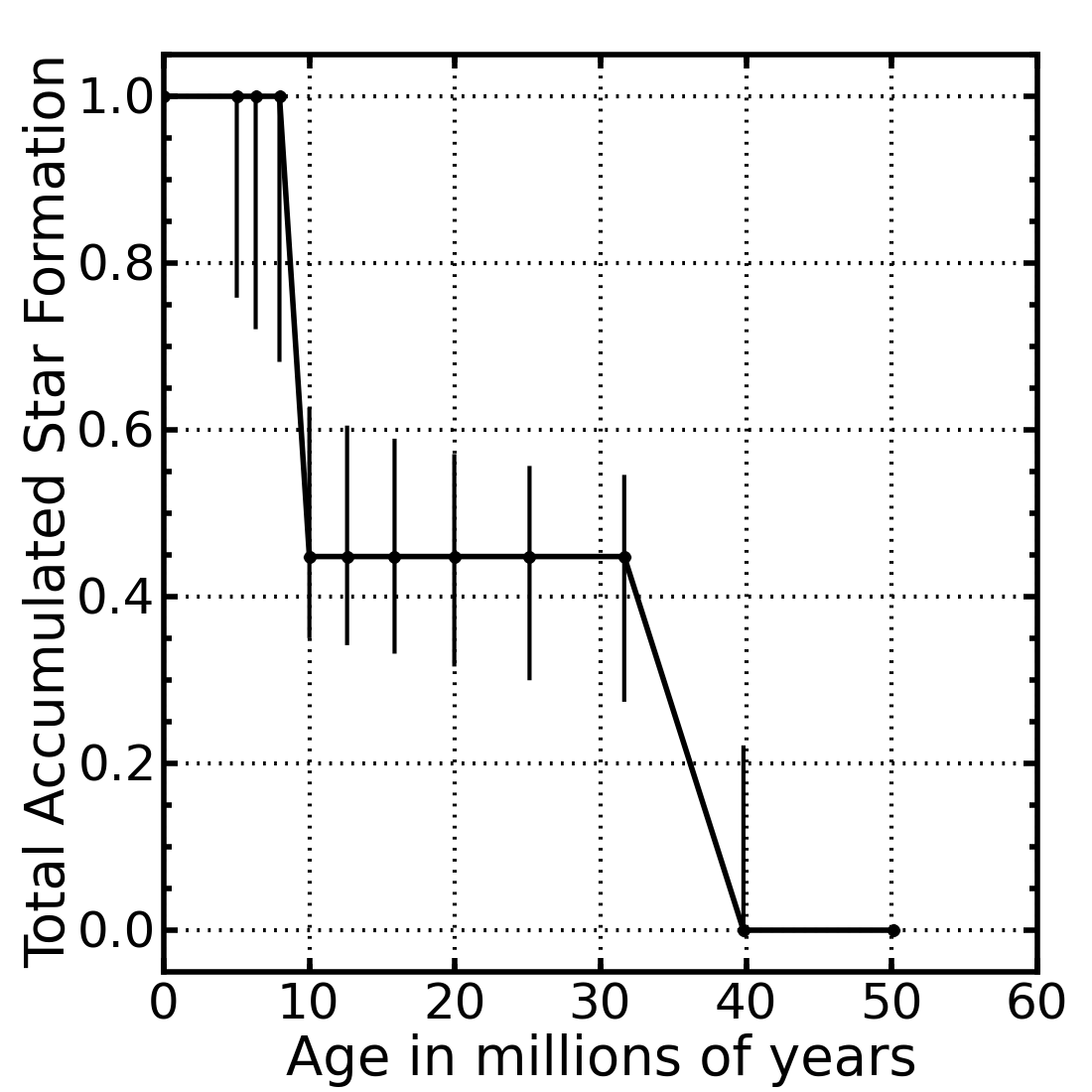}
\caption{1987A deep test.  {\it Upper-left} HST image of the region around SN1987A.  {\it Upper-right:}  Color-magnitude diagram from the HST WFPC2 imaging data of SN1987A.  The lines at F814W=18.5 and F814W=24 show the data we included in our sample for the shallow and deep fits, respectively.  {\it Lower-left:} Cumulative SFH resulting from the deep CMD fit.  With the full depth, the age is very tightly constrained.  {\it Lower-right:}  Cumulative SFH resulting from the shallow CMD fit.  The correct result is still preferred, but other ages are also allowed within the uncertainties.}
\label{87deep}
\end{figure}

An example of the efficacy of the method is in its application to SN
1987A.  We have run our model fits on deep WFPC2 photometry measured
from the archival data of proposal ID 7434 (PI: Kirshner).  We fit the
F555W-F814W CMD in the range 12$<$F814W$<$25 as shown in
Figure~\ref{87deep}, and get a well-constrained median
(22.0$^{+2.3}_{-5.8}$ M$_{\odot}$), which is consistent with the mass
(19$\pm 3$ M$_{\odot}$) derived in direct imaging studies
\citep{woosley88}, and with the combined mass of the binary merger
scenario
\citep[16$+$3$\rightarrow$19;][]{podsiadlowski1990,podsiadlowski1992}.
This comparison between techniques provides strong verification of our
proposed method.

While this test is encouraging, the data for 1987A is significantly
deeper than that of our more distant objects.  Two more tests,
however, suggest that our technique works even with much shallower
data.  First, we ran our model fits on 1987A only including the
photometry for stars brighter than apparent magnitude of 18.5
(absolute magnitude of 0), comparable to the depth of most of our more
distant targets.  The resulting median mass was more poorly
constrained (22.8$^{+2.5}_{-14.4}$ M$_{\odot}$), but was still
consistent with the known mass.  Thus, we may lose precision with
shallower data, but we can still obtain useful constraints.

In addition to our tests on SN 1987A, we have verified our technique
out to $\sim$8 Mpc by applying it to SN 2011dh in M51
\citep{murphy11a}, for which we found a progenitor mass of
13$^{+2}_{-1}$ M$_{\odot}$.  \citet{maund11} identified a progenitor
in archival HST images, which has since vanished \citep{vandyk2013},
and fit the bolometric luminosity to stellar-evolution models and
derived a progenitor mass of 13$\pm 3$ M$_{\odot}$, consistent with
our constraints.

\subsection{Resolved Stellar Photometry}

To generate the CMD, we measure resolved stellar photometry of the
{\it HST} field containing the location of the historic SN.  This
photometry was performed using the packages {\tt HSTPHOT} (for WFPC2
data) or {\tt DOLPHOT} \citep[for ACS data;]{dolphin2000}.  These
packages perform point spread function fitting optimized for the
undersampled flat-fielded images that come from {\it HST}.  All of the
photometry we use has been publicly released to the High-Level Science
Products in the HST archive through the ANGST and ANGRRR programs
(GO-10915 and AR-10945; PI: Dalcanton).  The details of the fitting
and culling parameters used are provided in \citet{dalcanton2009} and
the ANGRRR public data archive
\footnote{https://archive.stsci.edu/prepds/angrrr/}.  As part of these
programs, hundreds of thousands of artificial star tests were also
performed to assess completeness and photometric accuracy.  These
tests consist of inserting a single star into the data, rerunning the
the data reduction, and assessing whether the fake star was recovered,
and if so, how close its measured brightness was to the input
brightness.

\subsubsection{CMD Sample Selection}

To isolate the subset of stars from our catalogs that were co-spatial
with the historic supernovae, we used the coordinates for the SNe from
the Asiago Supernova Catalog \citep{barbon1999}, and galaxy distances
from \citet{dalcanton2009}.  We corrected the astrometry in our
catalogs by cross-correlating 2MASS positions for the bright stars in
our catalogs with our positions.  Our catalog astrometry is then
corrected such that the star positions agree with those of 2MASS as
precisely as our centroiding for these bright stars will allow
(typically $\sim$0.1$''$).  This correction to our catalog astrometry
made sure that our positions were at least as precise as those in the
SNe catalog (within a few tenths of an arcsecond).

With the location and the distance well-measured, we were able to pull
stars measured within a projected radius of 50~pc of each SNe.  In the
most distant cases, this radius is only a bit more than 1 arcsecond,
making the necessary precision of astrometry only $\sim$1$''$.

To provide fake star statistics for the photometric completeness and
precision appropriate to our sample, we required a minimum of ten
thousand fake stars into our images.  To reach this number, we
included fake stars from a region of up to a factor of 7 larger than
the real stars.  In fields where the quality of the data varies
quickly with position, such as near the center of M82, we applied
additional computing resources to obtain more artificial star tests
within the same radius as the stellar sample.  However, for almost all
of our fields, changes in stellar density, and therefore photometric
quality, were small over the field, making it possible to use a large
suite of artificial star tests to improve statistics on our CMD
fitting.

\subsubsection{CMD Fitting}

Our CMD fitting process was very similar to that performed in
\citet{jennings2012}.  We used the CMD-fitting package MATCH
\citep{dolphin2002,dolphin2013} to fit each CMD with the stellar
evolution models of \citet{girardi2002} with updates in
\citet{marigo2008} and \citet{girardi2010}.  The package allows models
to be shifted in temperature and luminosity space to mimic systematic
uncertainties, and it allows differential extinction to be applied to
the models during fitting.

\begin{figure}
\centerline{\includegraphics[width=4.0in]{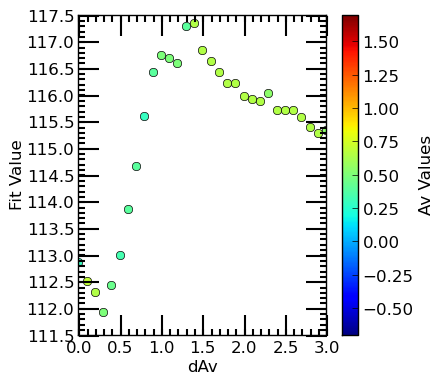}}
\caption{Technique for finding best-fit reddening and differential
extinction.  The best (lowest) value for the Poisson likelihood
statistic is compared for an extensive grid of A$_{\rm V}$, dA$_{\rm
V}$ combinations.  The best fit is then adopted as the SFH of the
sample.  This example is from SN2008iz, which had a best fit at
dA$_{\rm V}$=0.3, A$_{\rm V}$=0.3.}
\label{dav}
\end{figure}

First, we determined the best-fitting amount of differential
extinction to apply when fitting each SN.  We fitted the data with a
grid of values for the differential extinction ($dA_{\rm V}$) and
foreground extinction A$_{\rm V}$.  We chose the $dA_{\rm V}$ value
that provided the best fit to the data without requiring an $A_{\rm
  V}$ value below the known foreground extinction from
\citet{schlegel1998}.  We show an example plot summarizing this
extinction determination method in Figure~\ref{dav}.

\begin{figure}
\includegraphics[width=2.0in]{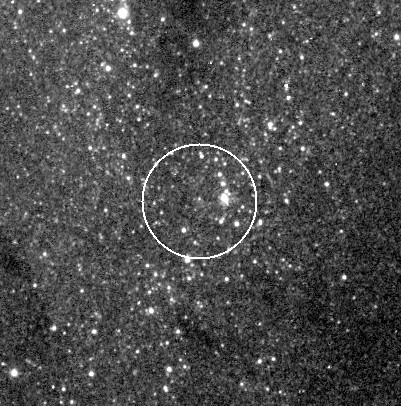}
\includegraphics[width=2.0in]{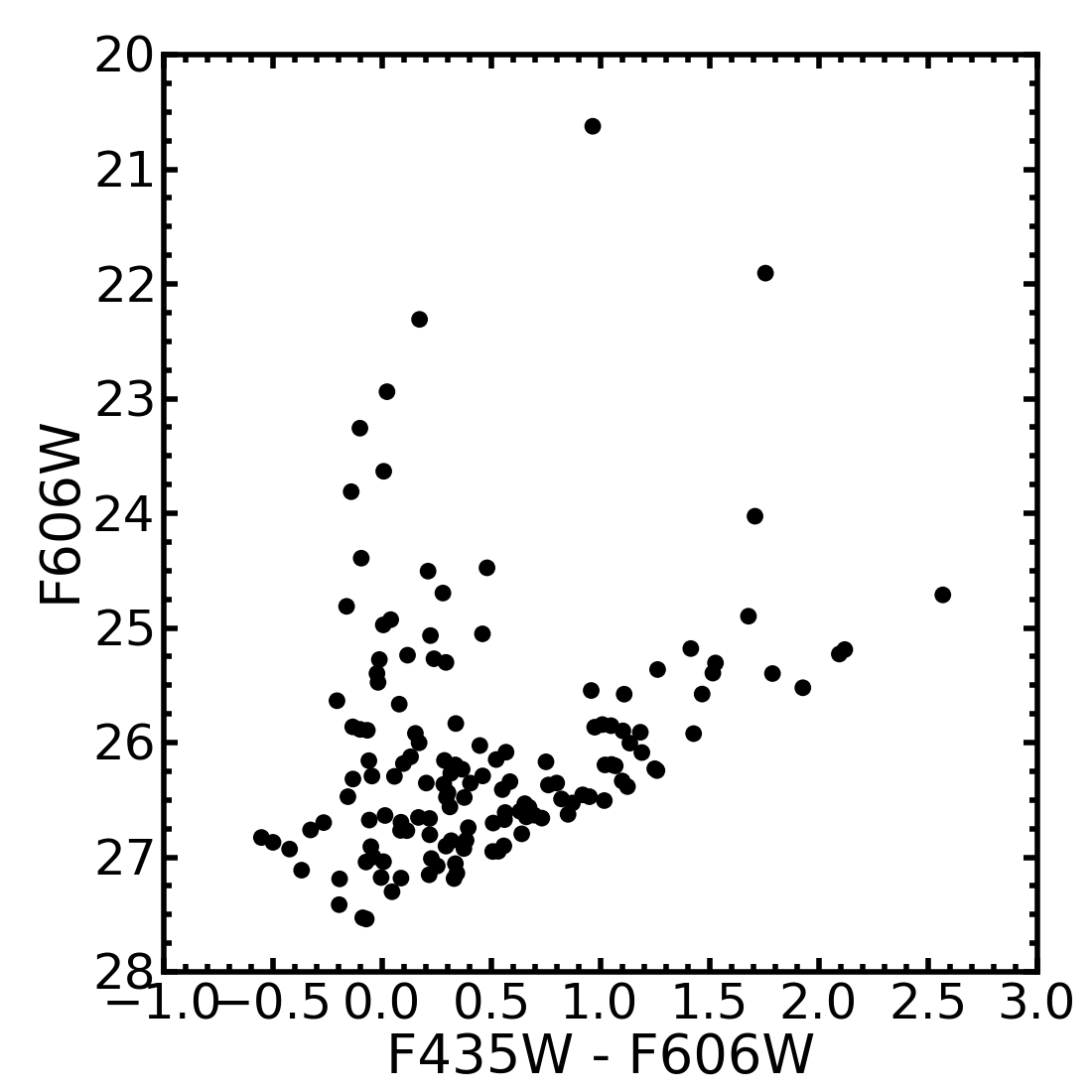}
\includegraphics[width=2.0in]{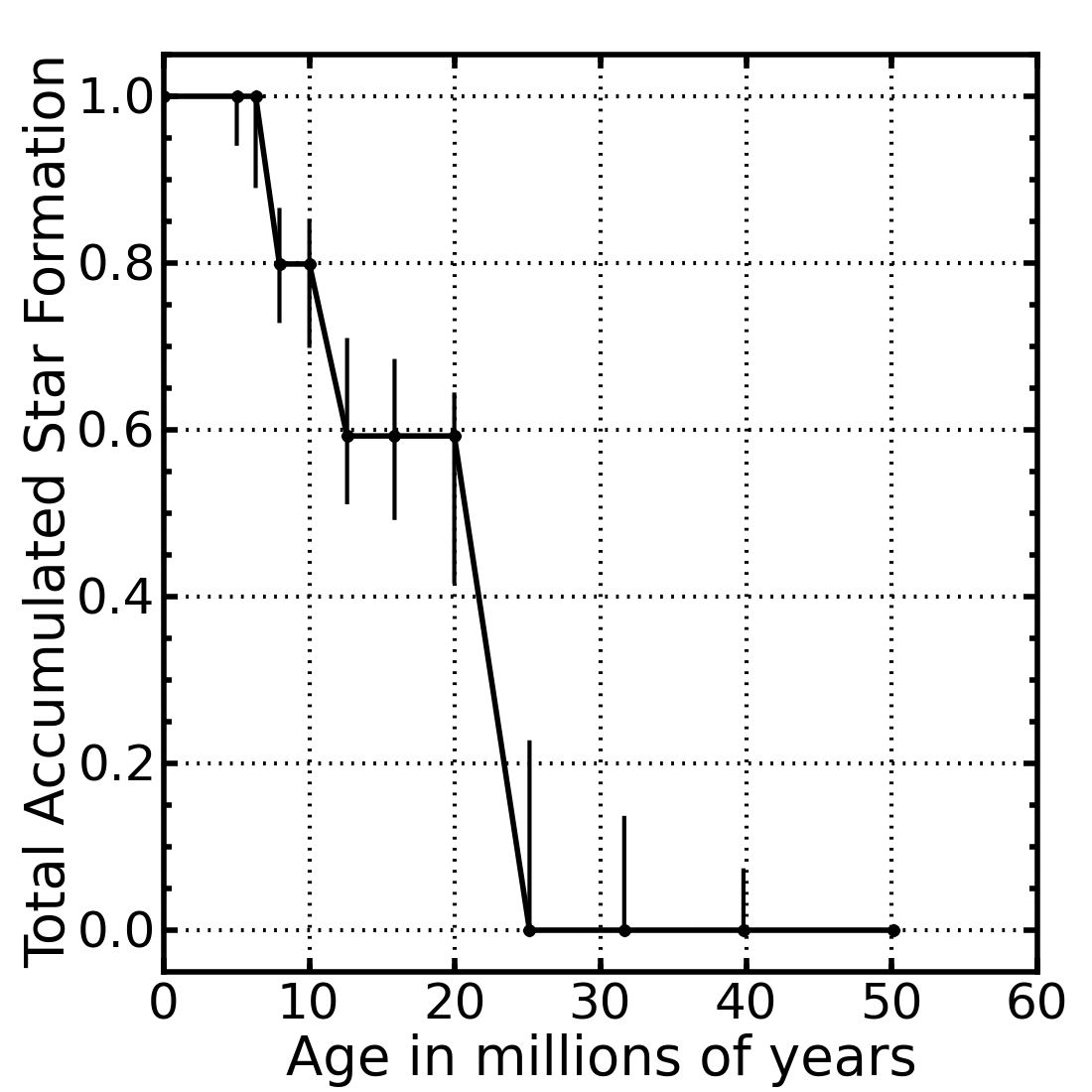}
\caption{Example of the results for a typical supernova location.  {\it Left:} 20$^{\prime\prime}{\times}20^{\prime\prime}$ image of region surrounding SN1993J in M81.  Our 50 pc extraction region is marked with the white circle.  {\it Center:} The color-magnitude diagram resulting from our photometry extraction. {\it Right} The cumulative fraction star formation from 50 Myr ago to the present as measured by our CMD-fitting analysis described in Section~2.}
\label{93J}
\end{figure}

\begin{figure}
\includegraphics[width=2.0in]{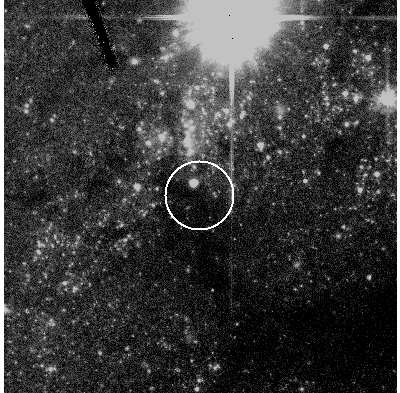}
\includegraphics[width=2.0in]{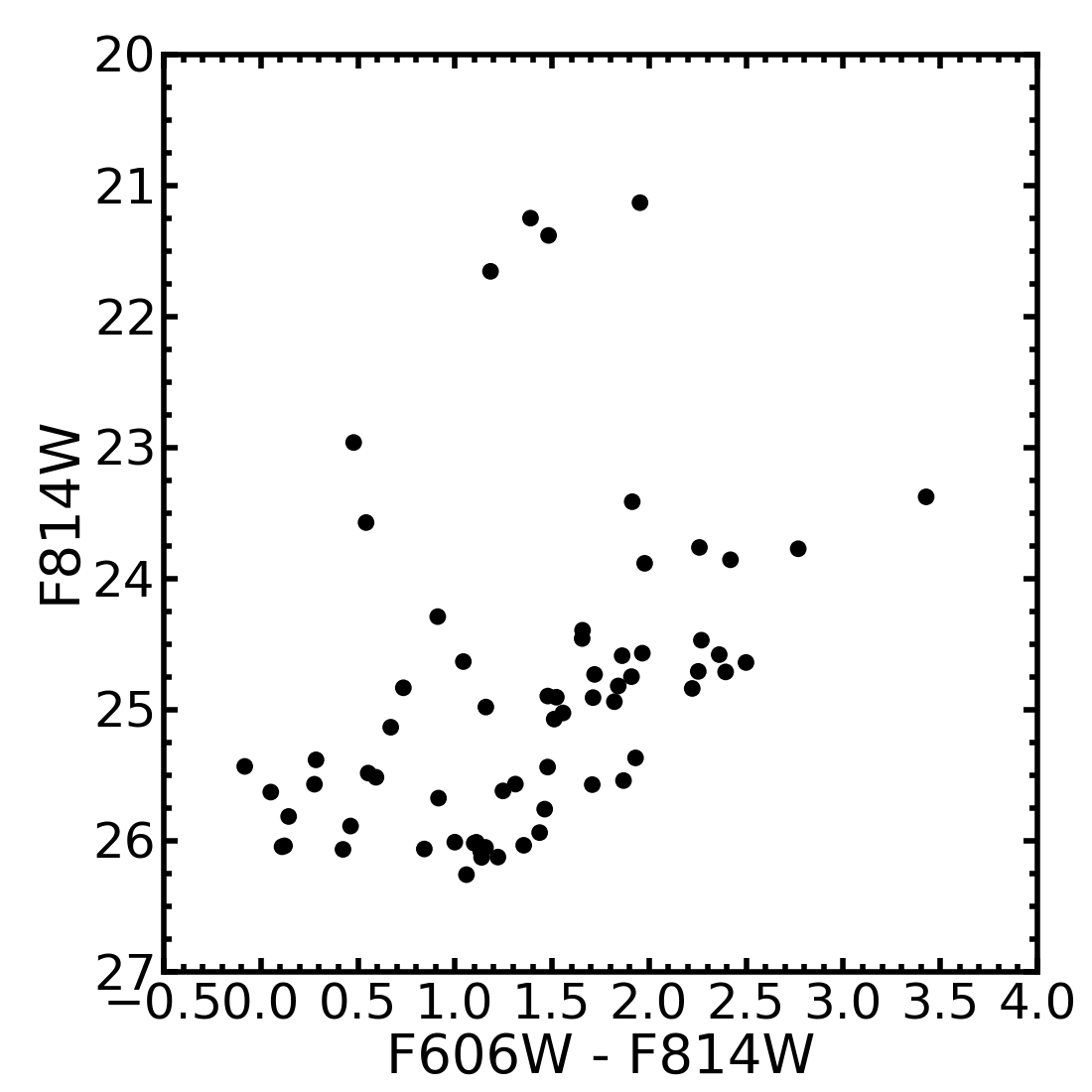}
\includegraphics[width=2.0in]{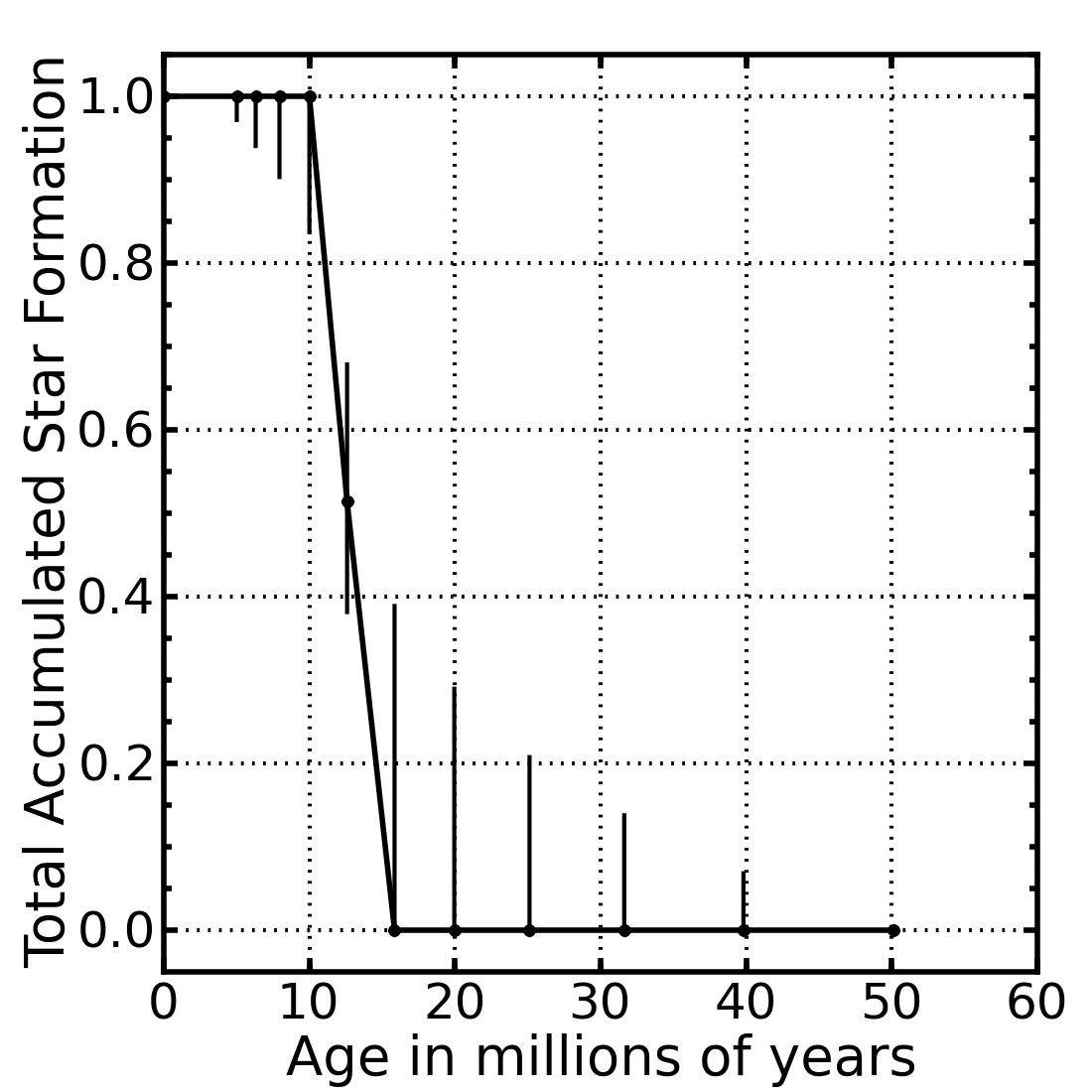}
\caption{Example of the results for a typical supernova location.  {\it Left:} 20$^{\prime\prime}{\times}20^{\prime\prime}$ image of region surrounding SN2002hh in NGC~6946.  Our 50 pc extraction region is marked with the white circle.  {\it Center:} The color-magnitude diagram resulting from our photometry extraction. {\it Right} The cumulative fraction star formation from 50 Myr ago to the present as measured by our CMD-fitting analysis described in Section~2.}
\label{02hh}
\end{figure}

\begin{figure}
\includegraphics[width=3.2in]{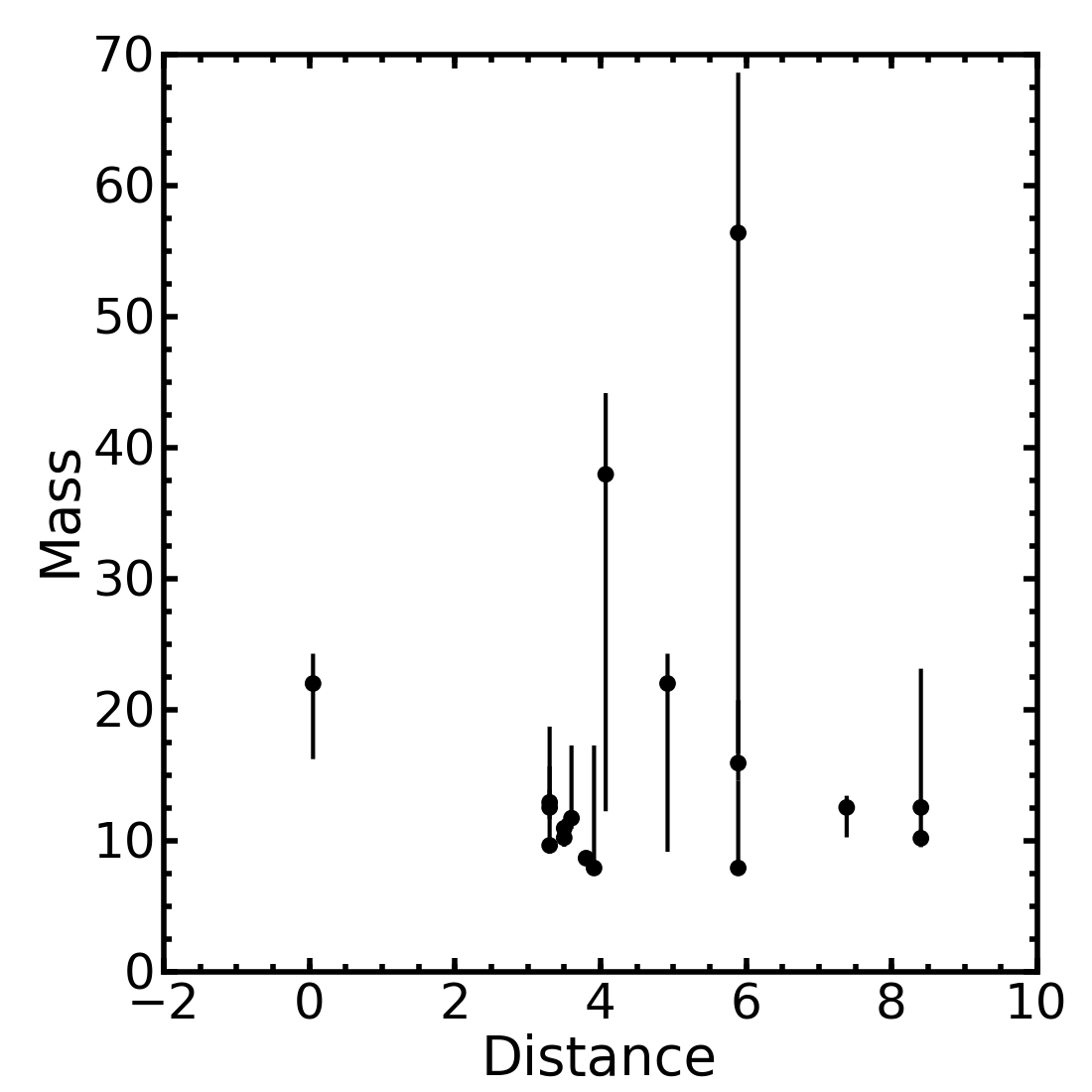}
\centerline{\hspace{-3.5in}\includegraphics[width=3.2in]{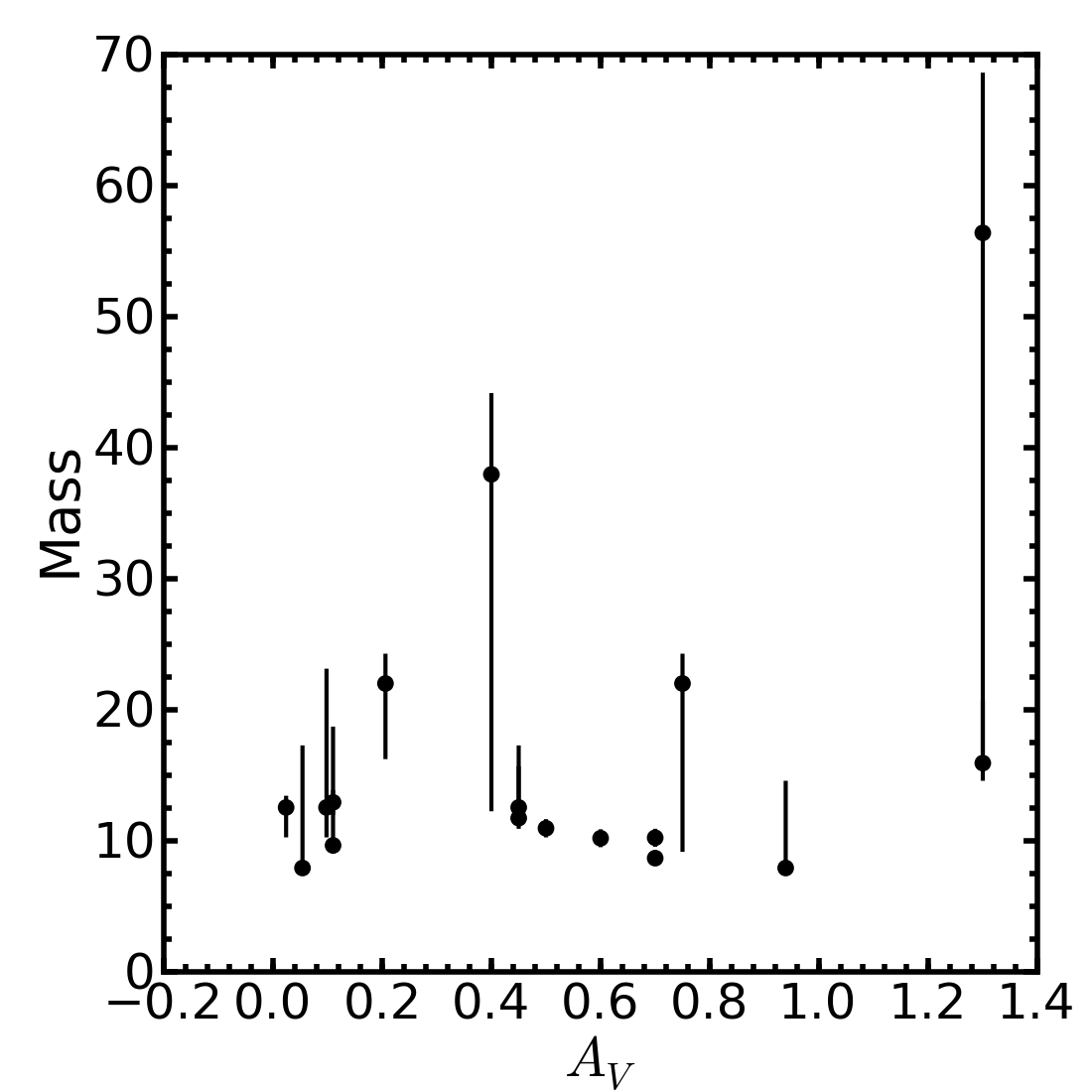}}
\caption{Scatter plots of our estimates of the progenitor masses
  against our adopted distance (left), and A$_{\rm V}$ (right).  No correlation
  is found, showing that these parameters did not significantly bias
  our results.}
\label{correlations1}
\end{figure}

With the distance, extinction, and differential extinction values
fixed, we fitted the CMD to find the most likely age distribution,
allowing metallicities for the young population in the range of
-0.6$\leq$[Fe/H]$\leq$0.1.  Examples for objects with data quality
typical of most of our sample are shown in
Figure~\ref{93J}-\ref{02hh}, where we show an image of our extraction
region, a plot of the CMD of the 50 pc region, and a final cumulative
star formation history from the fitting routine.  In
Figure~\ref{correlations1} we plot our final derived masses against
distance and $A_{\rm V}$.  The lack of correlations in the derived
masses as a function of these parameters suggests they do not
introduce any significant bias into our measurements.

To assess systematic uncertainties (due to any model deficiencies), we
reran the fitting with several changes to the models, following
\citet{dolphin2012}. We allowed the effective temperature of the
models to vary by $\Delta$log(T$_{eff}$)=0.02.  We allow the
bolometric luminosity of the models to vary by
$\Delta$log(L$_{bol}$)=0.17.  Furthermore, we allow the differential
extinction applied to the model to vary by ${\Delta}dA_{\rm V}$=0.2 in
cases of high $dA_{\rm V}$ ($>$0.4).  We run 100 fits to 100
realizations of the model, varying all of these parameters to account
for systematic uncertainties resulting from our stellar evolution
models and our treatment of the extinction.

Then, to measure the random uncertainties due to number of stars and
depth of the photometry, we use the {\tt hybridMC} task within the
MATCH package as described in \citet{dolphin2013}.  This task
determines the star formation rate that would allow an acceptable fit
to the data for each age bin, thus providing robust upper-limits for
bins where the best-fit star formation rates were 0.  With both
uncertainty determinations complete, we combine the random and
systematic uncertainties in quadrature using the MATCH routing {\it
  zcmerge} to calculate our final uncertainties on the star formation
rates in each age bin.

Next, we use our total uncertainties on the star formation rates in
each age bin to determine the uncertainty on the fraction of stellar
mass present in each age bin back to 50~Myr.  We perform 1000 Monte
Carlo realizations of the measured SFH from 50 Myr to the present.  We
then calculate the 16\% and 84\% ranges in stellar mass fraction in
each age bin from these tests.  We adopt these percentiles as the
uncertainties on fraction of stellar mass formed in each age bin
relative to the total stellar mass produced in the past 50 Myr.

\section{Progenitor Masses}

Once we have the mass fraction (and associated mass fraction
uncertainty) in each age bin, we calculate our first estimate of the
progenitor mass by determining the median age of the best fit.  We
then use our uncertainties to determine the age bins consistent with
containing the median to assign uncertainties on that median age.
Finally, we convert these ages to masses by taking the most massive
star remaining in the model isochrone corresponding to the each age
\citep[see][for more details]{jennings2012}.  These values provide our
the nominal progenitor mass and associated uncertainties for each SN.
These median masses and associated uncertainties ($\sigma_{med}$) are
provided in Table~\ref{median}.

Although the assignment of a single progenitor age is of interest,
many of our SFHs contain multiple coeval populations, making a more
complex distribution of the mass probability desirable for some
purposes.  We therefore have also tabulated the uncertainty for each
progenitor due to the spread in the recovered age distribution
($\sigma_{pop}$).  These uncertainties encompass 68\% of the total
population mass (about the median value of the best fit) with ages
$<$50 Myr including uncertainties and account for the full
distribution ages present at the SN location, similar to the technique
adopted in earlier work \citep{murphy11a,jennings2012}. In most cases,
the stellar mass is relatively well confined to a small age range, but
including this second set of uncertainties shows where there are
multiple ages present.  For example, the median age of the young
population surrounding SN1994I is well-determined, providing a
high-precision mass measurement of 10.2$\pm$0.7 M$_{\odot}$; however,
there is also a younger population present that represents a
significant fraction of the stellar mass.  If the presence of this
population is taken into account, the uncertainties on the progenitor
mass increase substantially to 10.2$^{+59.2}_{-1.8}$.  Thus, in this
case, only under the assumption that progenitor was a member of the
dominant young population is the mass of the progenitor
well-constrained.  Otherwise, it is only a lower limit.

Looking at these uncertainties, one can determine which SNe would
benefit most from improved photometry data. Large spreads in the 68\%
population mass accompanied by small errors on the median seem to
occur for SNe with few stars in the CMD.  For example, SN1951H has
only 11 stars for fitting, a 10\% error on the median, but
$\sigma_{pop}$ values that encompass the full mass range of the
models.  Thus, the well-constrained median suggests that the mass
should also be well constrained, but the small number of stars results
in large uncertainties for other age bins which would likely be
reduced with deeper data and a larger number of detected stars.

Finally, to provide detailed probability distributions for all SN, we
tabulate the probability that the progenitor was in each age/mass bin,
given the SFH and associated uncertainties.  These probability
distributions are given in Table~\ref{pdfs}, where each mass bin is
assigned a probability which comes from the most likely SFH, and an
associated uncertainty on the probability, which comes from applying
the uncertainties in the SFH to the mass probability distribution.
Thus, in order to account for both sources of uncertainty on the
progenitor mass (the fitting uncertainty and the uncertainty
associated with the intrinsic range of ages), it is necessary to
assign uncertainties to our probabilities.  However, in many cases,
the median mass is relatively well defined ($\sigma{<}20$\%), and
provides a simple, though less thorough, constraint on the progenitor
mass.

\begin{figure}
\centerline{\includegraphics[width=3.3in]{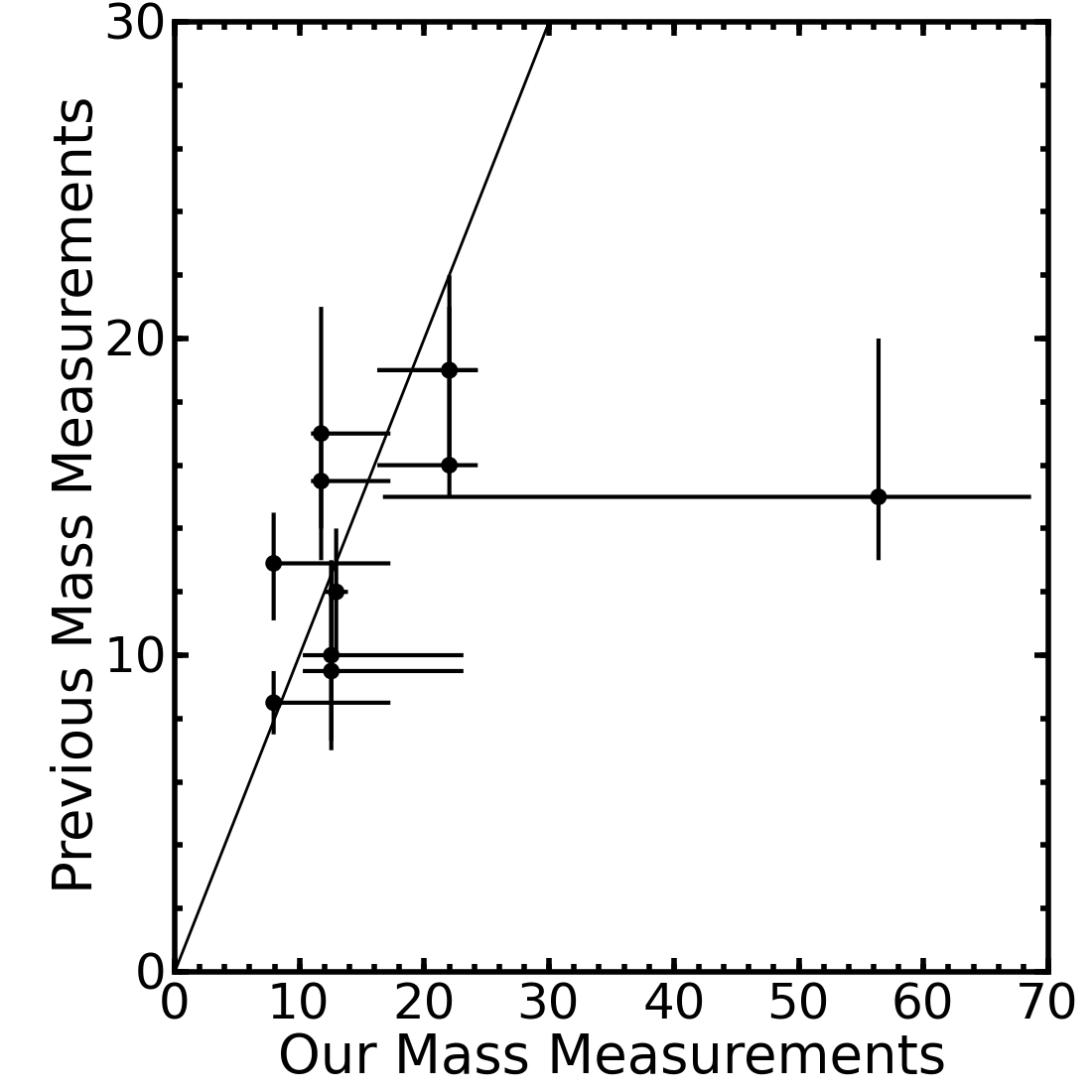}}
\caption{Our mass measurements plotted against all
  available measurements in the literature for the 6 SNe with
  literature measurements. The measurements are consistent within the
  uncertainties, with no systematic bias.  The large outlier is SN2004et, which was one of our least constrained sources due to the low number of stars and high differential extinction.}
\label{comp}
\end{figure}

Six of the events in our sample have previously-measured progenitor
masses from direct imaging (SN1987A, SN1993J, SN2004dj, SN2004et,
SN2005cs, SN2008bk, see Table~\ref{median}), and while our
measurements are less precise in some cases, they are consistent with
the previous measurements in all cases, as shown in Figure~\ref{comp}.
Indeed, in all cases the previous measurements are consistent with our
most optimistic uncertainties---the uncertainty on the median age of
the young population.

\subsection{Extreme Cases}

\begin{figure}
\includegraphics[width=2.0in]{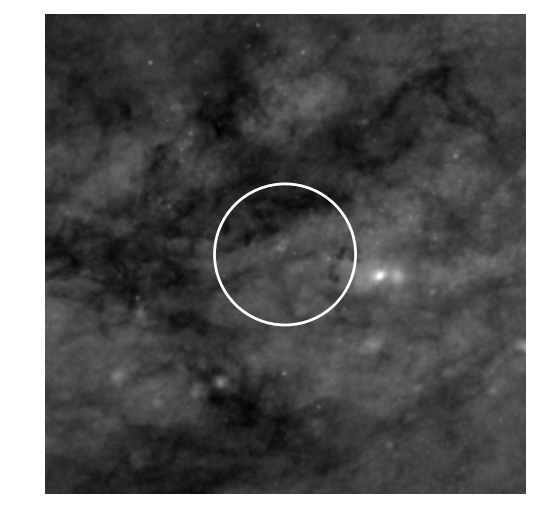}
\includegraphics[width=2.0in]{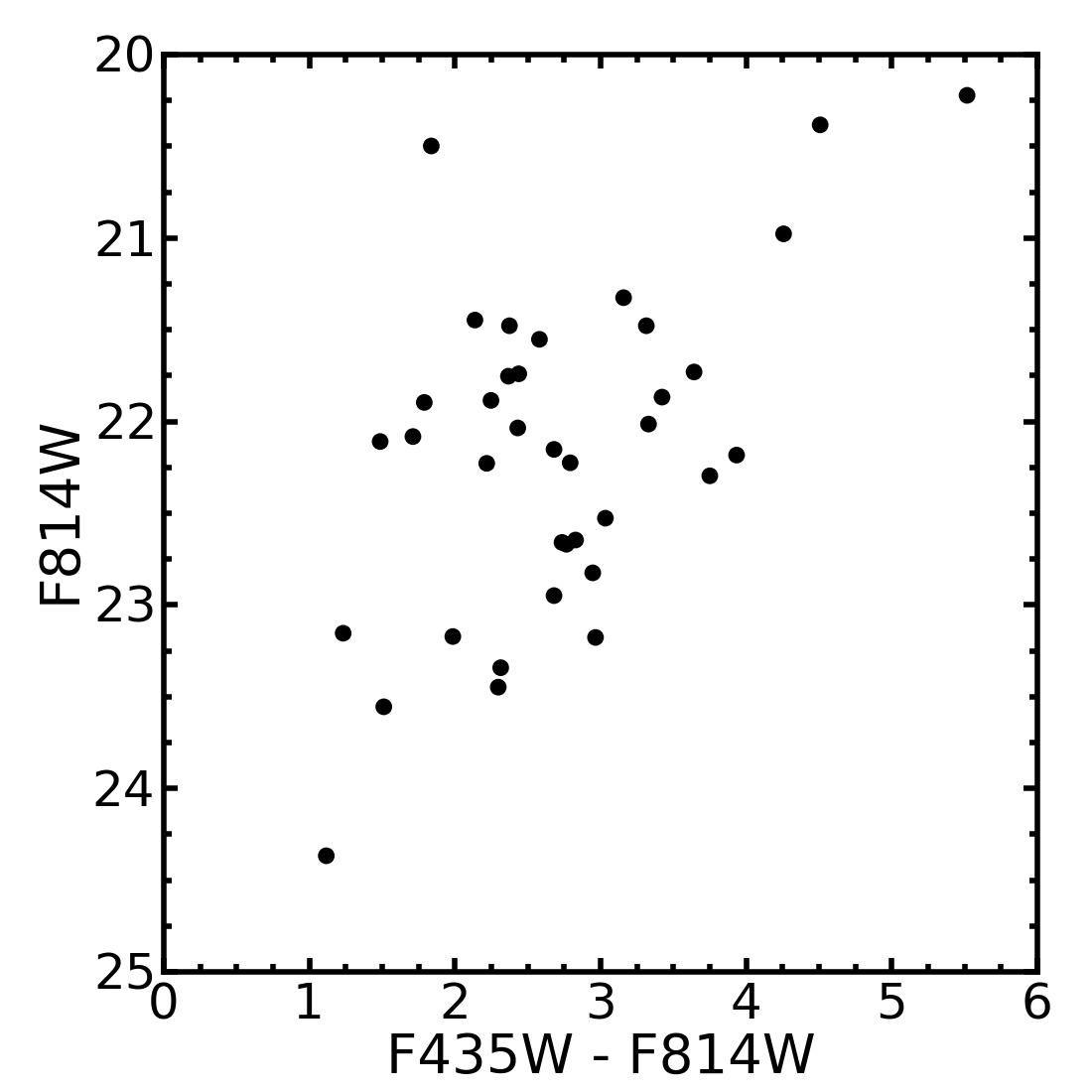}
\includegraphics[width=2.0in]{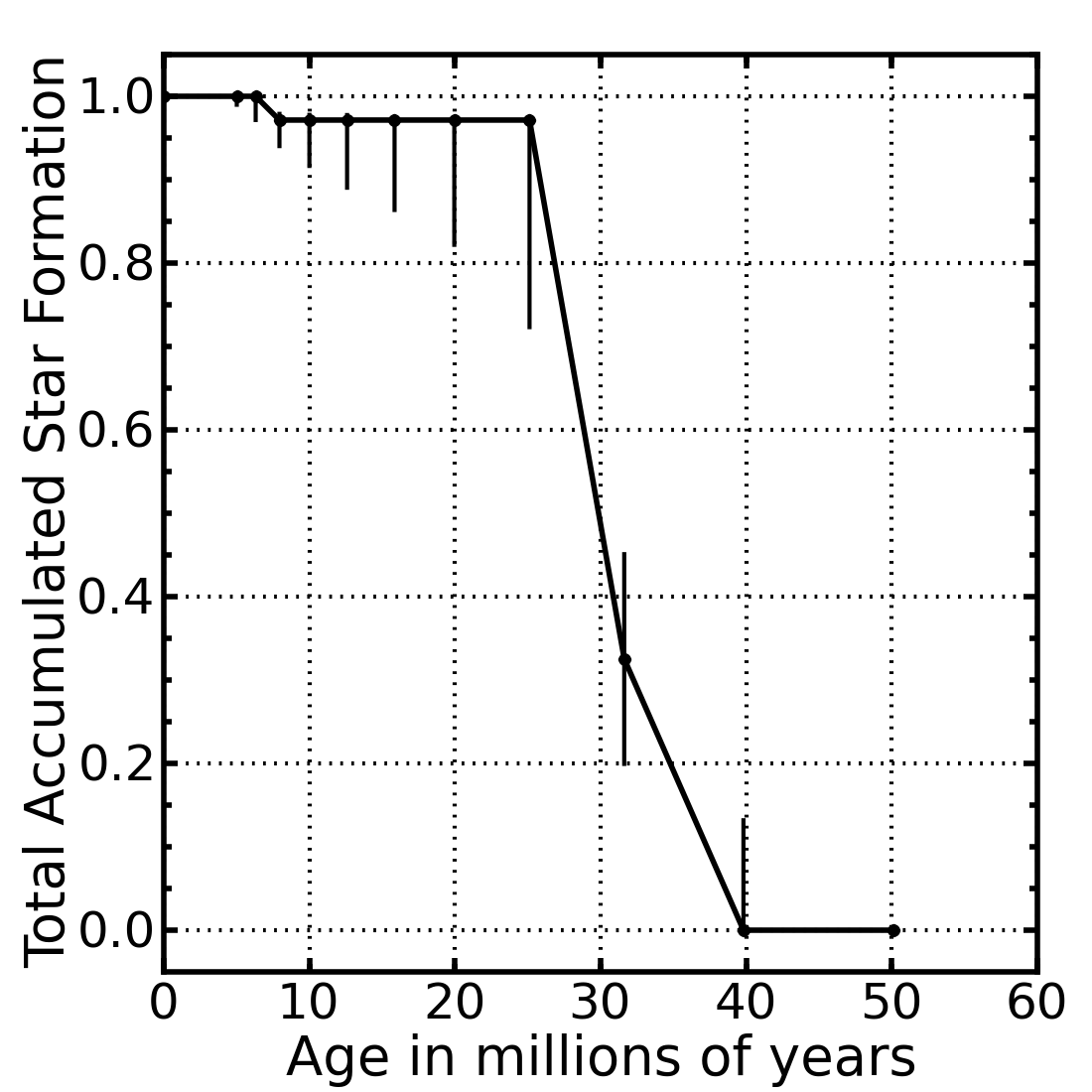}
\caption{Our measurement with the highest dust extinction.  {\it
Left:} 20$^{\prime\prime}{\times}20^{\prime\prime}$ image of region
surrounding SN2004am in M82.  Our 50 pc extraction region is marked
with the white circle.  {\it Center:} The color-magnitude diagram
resulting from our photometry extraction. The lack of any clear
feature suggests that our results are becoming less reliable at these
high amounts of differential extinction.  {\it Right:} The cumulative
fraction star formation from 50 Myr ago to the present as measured by
our CMD-fitting analysis described in Section~2.}
\label{04am}
\end{figure}

A few SNe in our samples stand out as extremely challenging of our
ability to measure star formation histories.  For example, in
Figure~\ref{04am} we show the image, photometry, and fitting results
for our most heavily extincted location, SN2004am.  In this case, even
though there is a very high amount of differential extinction
($dA_{\rm V}$=2.5), the relatively large number of stars provides a
good constraint on the age distribution.  Unfortunately, with this
much dust, there is clearly the possibility of a significant number of
more massive stars being completely hidden from the sample, which
would not be accounted for by our method.  We cannot account for stars
that are extincted out of our photometry sample.  Thus, this amount of
dust may make this result less reliable than many of the others.  Such
examples are unlikely to improve without much deeper data to probe to
very high extinctions.

\begin{figure}
\includegraphics[width=2.0in]{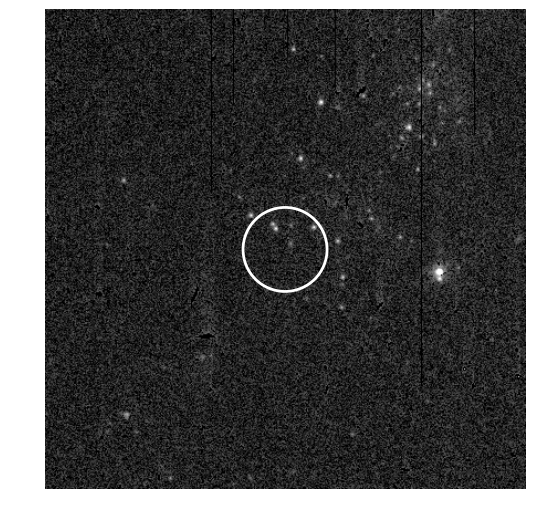}
\includegraphics[width=2.0in]{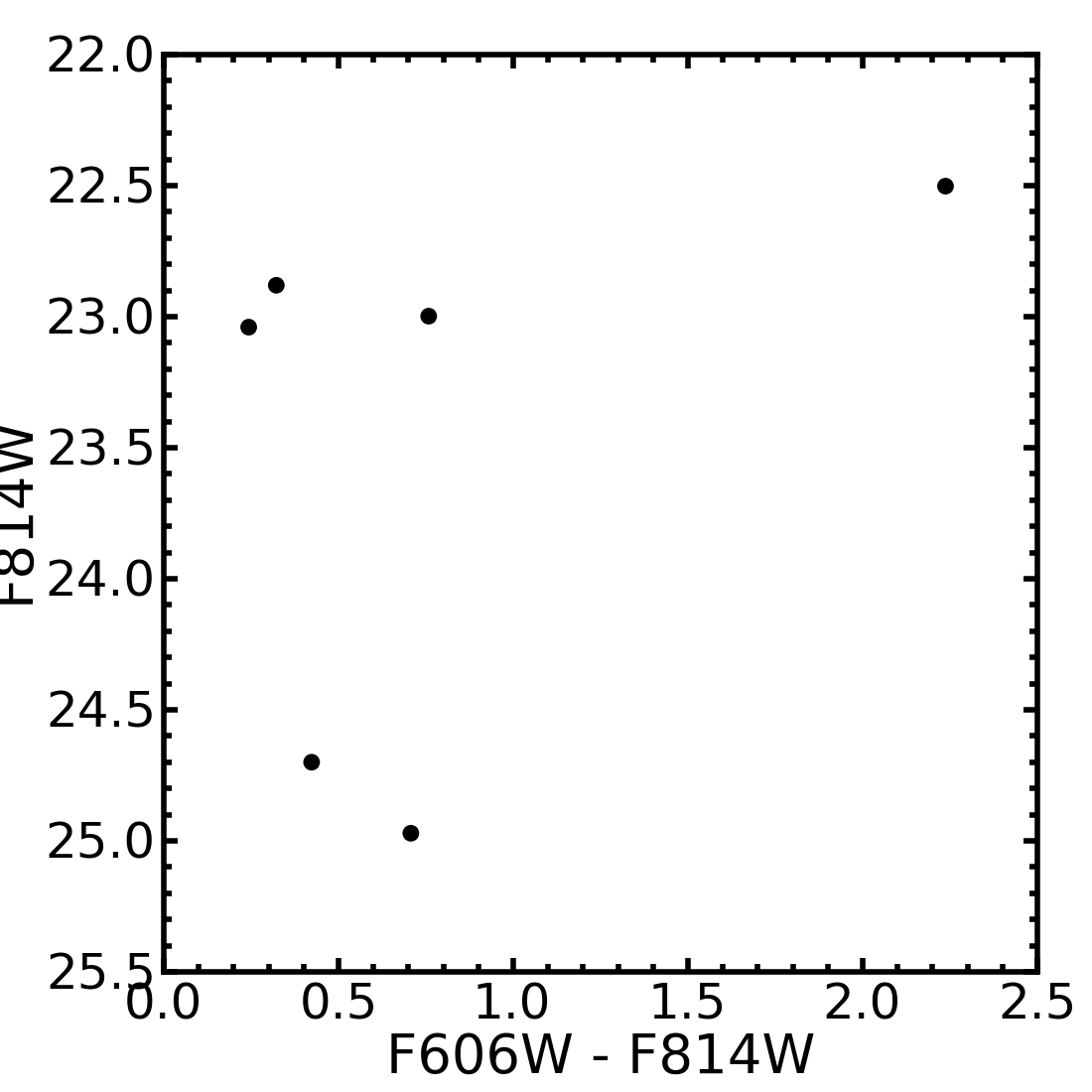}
\includegraphics[width=2.0in]{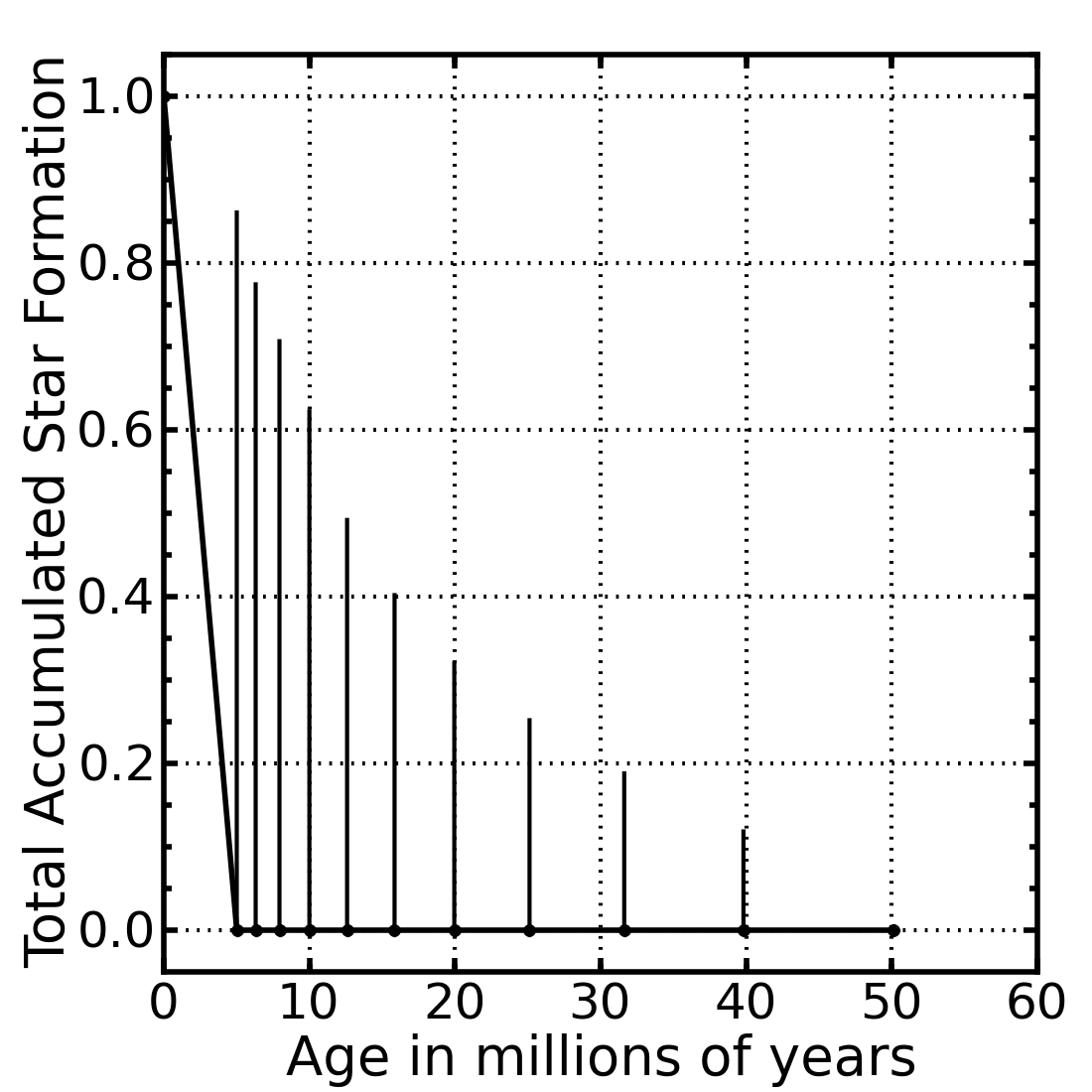}
\caption{Our measurement with the fewest number of stars.  {\it Left:}
20$^{\prime\prime}{\times}20^{\prime\prime}$ image of region
surrounding SN2004et in NGC6946.  Our 50 pc extraction region is
marked with the white circle.  {\it Center:} The color-magnitude
diagram resulting from our photometry extraction. {\it Right:} The
cumulative fraction star formation from 50 Myr ago to the present as
measured by our CMD-fitting analysis described in Section~2.  The low
precision makes it clear that we cannot constrain progenitor masses
with any fewer stars than this example.}
\label{04et}
\end{figure}

Another extreme case is SN2004et, where we only have 6 stars in the
CMD due to shallow imaging and a far distance (meaning a small
extraction region on the sky).  We show our results for this SN in
Figure~\ref{04et}, where the lack of stars results in very large
uncertainties.  Although the uncertainties on the progenitor mass are
large, the full range of masses are not allowed by our uncertainties,
which suggest the mass is $>$16 M$_{\odot}$.  These uncertainties are
reliable, as the best-mass is well away from (but within the large
errors of) the mass measured from direct imaging.  Interestingly, even
with the large uncertainties, the mass constraint is useful since it
rules out masses lower than the best-fit mass from direct imaging.
This example confirms that our uncertainty estimates are reliable, but
also demonstrates that attempting this technique with any fewer stars
is of little value.

Finally, our results in this work add further validation to our
method.  For the 6 SNe with we measure here that have literature
measurements, we plot our measurements against those from the
literature in Figure~\ref{comp}.  In all cases our measurements are
consistent with previous measurements within the uncertainties, and no
systematic bias is seen.

\subsection{Progenitor Mass Distribution}

\begin{figure}
\centerline{\includegraphics[width=3.3in]{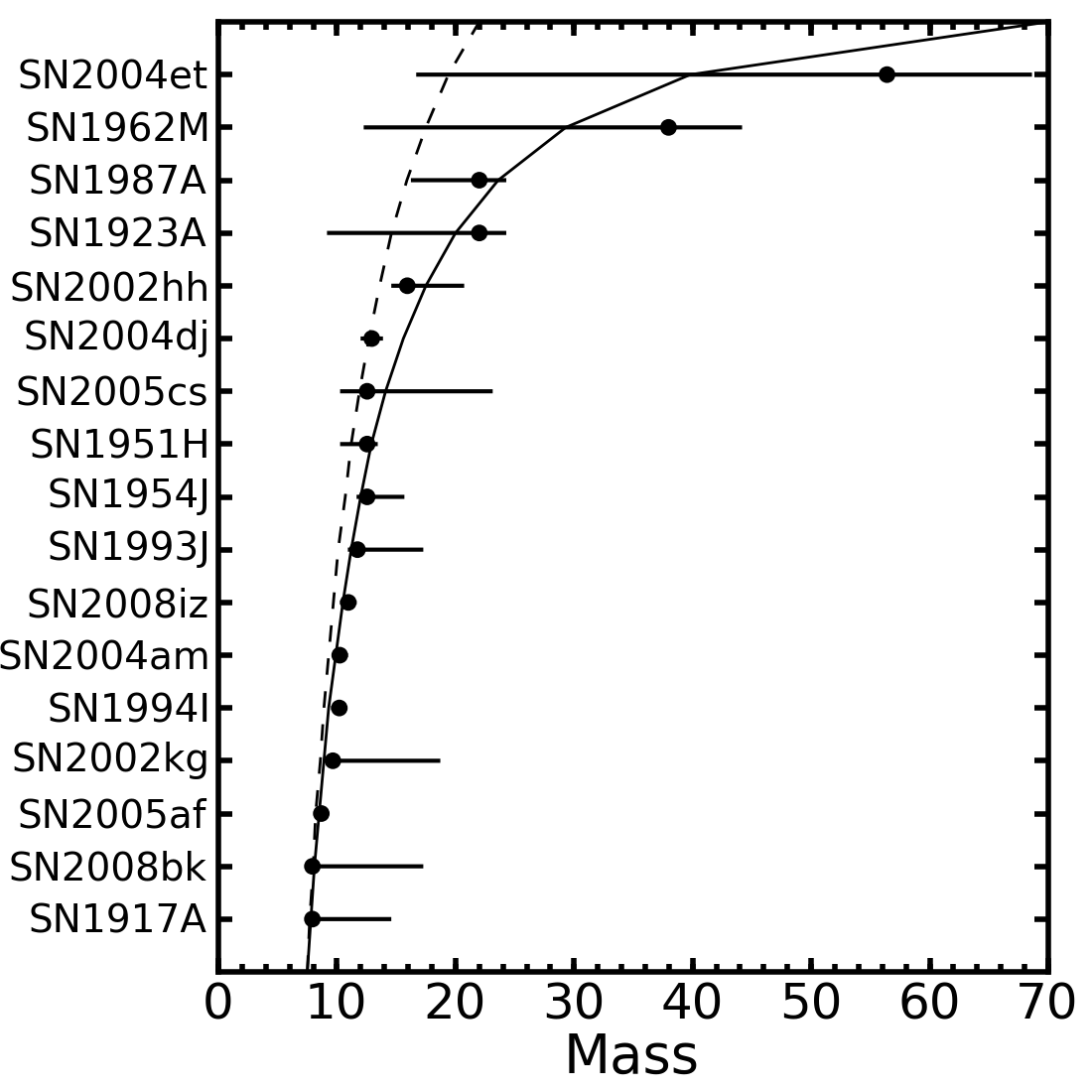}}
\caption{Ranked plot of our mass measurements and associated
uncertainties.  Overplotted are the expected mass distributions for a
\citet{salpeter1955} IMF with an upper-mass cutoff of 22~M$_{\odot}$
(dashed line), and with an upper mass cutoff of 70~M$_{\odot}$ (solid
line).}
\label{mass}
\end{figure}

We note that our results are consistent with no SN progenitors
$>$20~M$_{\odot}$, as are all of the progenitor mass measurements
currently available in the literature (see references in Section 1).
While we do have some best estimates that are higher mass, their
uncertainties all extend below 20~M$_{\odot}$.  Our most massive
central values are for SN2004et and SN1962M, but these only have 75\%
and 82\% probability of being $>$20~M$_{\odot}$.  Furthermore, the
direct imaging mass for SN2004et has an upper limit of 20~M$_{\odot}$,
suggesting that the correct mass is indeed at the low end of our
uncertainties.  Figure~\ref{mass} plots the masses in ranked order,
along with the expected distribution of masses for a
\citet{salpeter1955} IMF with different upper-mass cutoffs.  The large
uncertainties on the high progenitor masses severely limit our ability
to determine the existence of such a cutoff.  Thus, our current sample
and data quality does not provide any conclusive evidence that
high-mass stars produce core-collapse supernovae.  This lack of
conclusive $>$20~M$_{\odot}$ progenitors is consistent with findings
of several other studies \citep{smartt09a,jennings2012}, hinting that
there could be a ceiling to SN production or a mass range that
under-produces SNe.  However, if we can measure a single progenitor
mass $>$20~M$_{\odot}$ with even 20\% precision, constraints on the
progenitor mass distribution would be greatly improved.

\section{Conclusions}

We have constrained the progenitor masses of 17 historic SNe using CMD
fitting of stellar populations measured from HST archival data.
Eleven of these are new constraints, making the total number of
historic SN progenitor masses 28.  Even with this dramatic increase in
mass measurements, there is still not a single high-precision
measurement of a progenitor $>$20~M$_{\odot}$, making characterization
of the progenitor mass distribution difficult.

This work represents all that is possible with the current state of
the HST archive.  The power of the technique is clear, and we hope
that future studies will be made possible by more and deeper HST
imaging of nearby galaxies containing historic SNe.

Support for this work was provided by NASA through grants AR-13277,
GO-10915, and Hubble Fellowship grant 51273.01 from the Space
Telescope Science Institute, which is operated by the Association of
Universities for Research in Astronomy, Inc., for NASA, under contract
NAS 5-26555. Z.G.J. is supported in part by a National Science
Foundation Graduate Research Fellowship.


\clearpage

\bigskip

\begin{turnpage}

\begin{deluxetable}{lcccccccccc} 
\tablewidth{8in}
\tablecaption{SN sample\tablenotemark{a}}
\tablehead{
\colhead{SN} &
\colhead{Type} &
\colhead{RA}  &
\colhead{DEC}  &
\colhead{Galaxy} &
\colhead{PID} &
\colhead{m-M} &
\colhead{Mpc} &
\colhead{$A_V$} &
\colhead{d$A_V$} &
\colhead{N$_{stars}$}
}
\startdata
SN1917A  & II  &  20:34:46.90  & 60:07:29.00  &  NGC 6946  & 11229      & 28.9 & 6.0 & 0.939 &  0.0   & 7		       \\
SN1923A	 & II P  &  13:37:09.20  & -29:51:04.00 &  NGC 5236 (M83) & 11360 & 28.5 & 5.0 &0.75 &   1.6   & 142		       \\
SN1951H	 & II  &  14:03:55.30  & 54:21:41.00  &  NGC 5457 (M101) & 9490 & 29.3 & 7.2 & 0.024 &  0.0    & 11		       \\
SN1954J	& II pec   &  07:36:55.36  & 65:37:52.1   &  NGC 2403	 & 10182    &   28.0 & 4.0 & 0.45 &   0.2  & 212		       \\
SN1962M	 & II P  &  03:18:12.20  & -66:31:38.00 &  NGC 1313	&9796     &   28.0 & 4.0 & 0.4 &    0.0  & 125		       \\
SN1980K	& II L   &  20:35:30.07  & 60:06:23.75  &  NGC 6946 & 11229	     &   28.9 & 6.0 &\nodata & \nodata & 2		       \\
SN1985F	 & Ib  &  12:41:33.01  & 41:09:05.94  &  NGC 4618 & 9042	     &   29.5 & 7.9 & \nodata & \nodata & 4		       \\
SN1987A	 & II pec  &  05:35:28.01  & -69:16:11.61 &  LMC & 7434	     &   18.5 & 0.05 & 0.206 &  0.7  & 11800		       \\
SN1993J	& IIb   &  09:55:24.95  & 69:01:13.38  &  NGC 3031 (M81) & 10584  & 28.0 & 4.0 & 0.45 &   0.0   & 143		       \\
SN1994I	 & Ic  &  13:29:54.07  & 47:11:30.50  &  NGC 5194	& 10452     &   29.6 & 8.3 & 0.6  &   0.0  & 42		       \\
SN2002ap & Ic pec  &  01:36:23.85  & 15:45:13.20  &  NGC 0628	& 10272     &   29.3 & 7.2 & \nodata & \nodata  & 1		       \\
SN2002bu & IIn  &  12:17:37.18  & 45:38:47.40  &  NGC 4242	& 10272     &   29.5 & 7.9 &\nodata & \nodata  & 0		       \\
SN2002hh & II P  &  20:34:44.29  & 60:07:19.00  &  NGC 6946 & 10607	     &   28.9 & 6.0 &  1.3  &   0.0 & 66		       \\
SN2002kg & IIn  &  07:37:01.83  & 65:34:29.30  &  NGC 2403  & 10182	     &   28.0 & 4.0 & 0.11 &   0.0  & 267		       \\
SN2003gd & Ic pec  &  01:36:42.65  & 15:44:19.90  &  NGC 0628  & 10272	     &   29.3 & 7.2 & \nodata & \nodata  & 4		       \\
SN2004am & II P  &  09:55:46.61  & 69:40:38.10  &  NGC 3034 (M82) & 10776  & 28.0 & 4.0 &0.7 &    2.5    & 37		       \\
SN2004dj & II P  &  07:37:17.02  & 65:35:57.80  &  NGC 2403 & 10607	     &   28.0 & 4.0 &  0.11&    0.0  & 127		       \\
SN2004et & II P  &  20:35:25.33  & 60:07:17.70  &  NGC 6946  & 11229	     &   28.9 & 6.0 &1.3  &   1.4   & 6		       \\
SN2005af & II  &  13:04:44.06  & -49:33:59.80 &  NGC 4945  & 10877	     &   27.9 & 3.8 &0.7  &   0.0   & 6		       \\
SN2005cs & II P  &  13:29:52.85  & 47:10:36.30  &  NGC 5194  & 10452	     &   29.6 & 8.3 & 0.098 &  0.0   & 33		       \\
SN2008bk & II P  &  23:57:50.42  & -32:33:21.50 &  NGC 7793 & 12196	     &   28.0 & 4.0 & 0.054 &  0.0   & 248		       \\
SN2008iz & II  &  09:55:51.55  & 69:40:45.80  &  NGC 3034 (M82) & 10776  & 28.0 & 4.0 & 0.5  &   0.3    & 56                     \\ 
\enddata
\tablenotetext{a}{Columns are (1) Name of SN, (2) Spectral Type of the SN, (3) Right Ascension of the SN in J2000 coordinates, (4) Declination of the SN in J2000 coordinates, (5) host galaxy of the SN, (6) program identification number of the HST data used for our photometry, (7) Distance modulus to the galaxy in magnitudes, (8) Distance to the galaxy in Mpc, (9) Foreground extinction to the SN in magnitudes, (10) differential extinction applied to our model fits in magnitudes, and (11) the number of stars in our photometry within the physical radius of 50 pc.} 
\label{sample}
\end{deluxetable}

\begin{deluxetable}{lcccccc}
\tablewidth{8in}
\tablecaption{Median mass, uncertainties, and 1$\sigma$ mass distributions\tablenotemark{a}}
\tablehead{
\colhead{SN} &
\colhead{Median Mass}  &
\colhead{$+\sigma_{med}$}  &
\colhead{$-\sigma_{med}$} &
\colhead{$+\sigma_{pop}$} &
\colhead{$-\sigma_{pop}$} &
\colhead{Direct Imaging Mass}}
\startdata
SN1917A  &   7.9  &   6.7      &       0.5      &       25.1   &     0.5 & \nodata\\
SN1923A  &   22.0  &  2.3      &       12.8      &      45.6    &    13.8 & \nodata\\
SN1951H  &   12.5  &  0.9      &       2.3      &       6.2     &    5.2 & \nodata\\
SN1954J  &   12.5  &  3.1       &      0.9       &      16.7    &    4.4 & \nodata\\
SN1962M  &   38.0  &  6.2      &       25.7      &      6.2     &    30.6 & \nodata\\ 
SN1987A (shallow)  &   22.8 &   2.5     &        14.4       &     47.4    &    15.5 & 19$\pm$3 \citep{woosley88}\\
SN1987A (deep)  &   22.0  &  2.3      &       5.8       &      8.6     &    13.8 & 19$\pm$3 \citep{woosley88}\\
  &   &    &        &       &   & Binary merger 16$+$3$\rightarrow$19\\
  &   &    &        &       &   &  \citep{podsiadlowski1990,podsiadlowski1992}\\
SN1993J   &  11.7  &  5.5      &       0.8       &      21.3    &    2.2 & $\sim$17 \citep{aldering1994}\\
 &   &     &       &     &   & Two stars, each 14--17\\
 &   &     &       &     &   & \citep{stancliffe2009}\\
SN1994I   &  10.2  &  0.7      &       0.7       &      59.2    &    1.8 & \nodata\\
SN2002hh  &  15.9  &  4.8      &       1.3       &      9.9     &    6.4 & \nodata\\
SN2002kg  &  9.7   &  9.1      &       0.6       &      13.5     &   2.3 & \nodata\\
SN2004am  &  10.2  &  0.7      &       0.7       &      4.3     &    1.8 & \nodata\\
SN2004dj  &  12.9  &  0.9      &       0.9       &      17.7    &    4.7 & $\sim$12 \citep{wang2005}\\
SN2004et  &  56.4  &  12.2     &       39.7      &      12.2    &    48.1 & 15$^{+5}_{-2}$ \citep{li2005}\\
SN2005af  &  8.7   &  0.5      &       0.5       &      5.2     &    1.3 & \nodata\\
SN2005cs  &  12.5  &  10.6    &        2.3       &      54.0    &    4.4 & 10$\pm$3 \citep{li2006}\\
 &   &     &       &     &   & 9.5$^{+3.4}_{-2.2}$ \citep{maund2014b}\\
SN2008bk  &  7.9   &  9.4     &        0.5       &      25.1    &    0.5 & 12.9$^{+1.6}_{-1.8}$ \citep{maund2014}\\
  &   &    &        &       &   & 8.5$\pm$1 \citep{mattila2008}\\
SN2008iz  &  11.0  &  0.7      &       0.7        &     4.7     &    2.8 & \nodata\\
\enddata
\label{median}
\tablenotetext{a}{Columns are (1) Name of SN, (2) Mass in solar units corresponding to the median age of the stars $<$50 Myr old within 50 pc of the event, (3) positive 1$\sigma$ uncertainty on the median mass in solar units, (4) negative 1$\sigma$ uncertainty on the median mass in solar units, (5) positive 1$\sigma$ uncertainty on the mass including the entire $<$50 Myr age distribution of the surrounding population, (6) negative 1$\sigma$ uncertainty on the mass including the entire $<$50 Myr age distribution of the surrounding population, and (7) the measurement of the progenitor mass from direct imaging.}
\end{deluxetable}

\end{turnpage}

\clearpage

\clearpage

\clearpage


\begin{deluxetable}{lccccc}
\tablecaption{Probability distributions, with associated uncertainties, for our sample,\tablenotemark{a}~~full table available in machine readable format only.}
\tablehead{
\colhead{SN} &
\colhead{Mass High}  &
\colhead{Mass Low}  &
\colhead{Probability} &
\colhead{$+$uncertainty}  &
\colhead{$-$uncertainty}}
\startdata
SN1917A  &  68.6  &  45.3  &  0.0  &  5.7  &  0.0 \\
SN1917A  &  45.3  &  33.0  &  0.0  &  11.0  &  0.0 \\
SN1917A  &  33.0  &  25.9  &  10.6  &  9.9  &  10.6 \\
SN1917A  &  25.9  &  20.7  &  0.0  &  20.7  &  0.0 \\
SN1917A  &  20.7  &  17.3  &  0.0  &  27.2  &  0.0 \\
SN1917A  &  17.3  &  14.6  &  0.0  &  34.7  &  0.0 \\
SN1917A  &  14.6  &  12.5  &  0.0  &  42.4  &  0.0 \\
SN1917A  &  12.5  &  10.9  &  0.0  &  50.4  &  0.0 \\
SN1917A  &  10.9  &  9.6  &  0.0  &  59.2  &  0.0 \\
SN1917A  &  9.6  &  8.4  &  0.0  &  73.1  &  0.0 \\
...  &  ...  &  ...  &  ...  &  ...  &  ... \\
\enddata
\label{pdfs}
\tablenotetext{a}{Columns are (1) Name of SN, (2) upper mass limit of the mass bin, (3) lower mass limit of the mass bin, (4) most likely percentage of stellar mass ($<$50 Myr) in the mass bin, (5) positive 1$\sigma$ uncertainty on the percentage, and (6) negative  1$\sigma$ uncertainty on the percentage.}
\end{deluxetable}

\end{document}